\documentclass[12pt]{phb-proc4-auth}

\usepackage{graphicx}
\usepackage{amssymb}

\begin{document}

\begin{frontmatter}

\journal{SNS'2004}

\title{Anisotropic Electron-Phonon Coupling Uncovered By Angle-Resolved Photoemission}

\author[CA]{T. P. Devereaux\corauthref{1},}
\author[US]{T. Cuk,}
\author[US]{Z.-X. Shen,}
\author[JP]{and N. Nagaosa}

\address[CA]{Department of Physics, University of Waterloo, Waterloo, Ontario, N2L
3G1, Canada}
\address[US]{Dept. of Physics, Applied Physics and Stanford
Synchrotron Radiation Laboratory, Stanford University, California
94305, USA}
\address[JP]{Department of Applied Physics, University of Tokyo,
Bunkyo-ku, Tokyo 113-8656, Japan}

\corauth[1]{Corresponding Author Email: tpd@lorax.uwaterloo.ca}

\begin{abstract}
Recently there has been an accumulation of experimental evidence in
the high temperature superconductors suggesting the relevance of
electron-phonon coupling in these materials. These findings
challenge some well-held beliefs of what electron-phonon
interactions can and cannot do. In this article we review evidence
primarily from angle-resolved photoemission (ARPES) measurements
which point out the importance of electronic coupling to certain
phonon modes in the cuprates.
\end{abstract}


\end{frontmatter}

The physics underlying the peculiar behavior of the high
temperature superconductors in the normal state and the high
transition temperatures themselves is far from understood. While
the importance of magnetic interactions has been widely pointed
out, a number of experimental findings strongly suggest that
electron-phonon coupling must play a role in these materials.
These findings include strong phonon renormalizations with
temperature and/or doping seen in neutron and Raman
scattering\cite{B1g,breathing}, the doping dependence of the
isotope effect of the transition temperature\cite{isotope}, the
isotope dependence of the superfluid density\cite{Keller}, and
lastly the pressure, layer and material dependence of T$_{c}$
itself.

\begin{figure*}[htb]
\includegraphics[width= 1\textwidth]{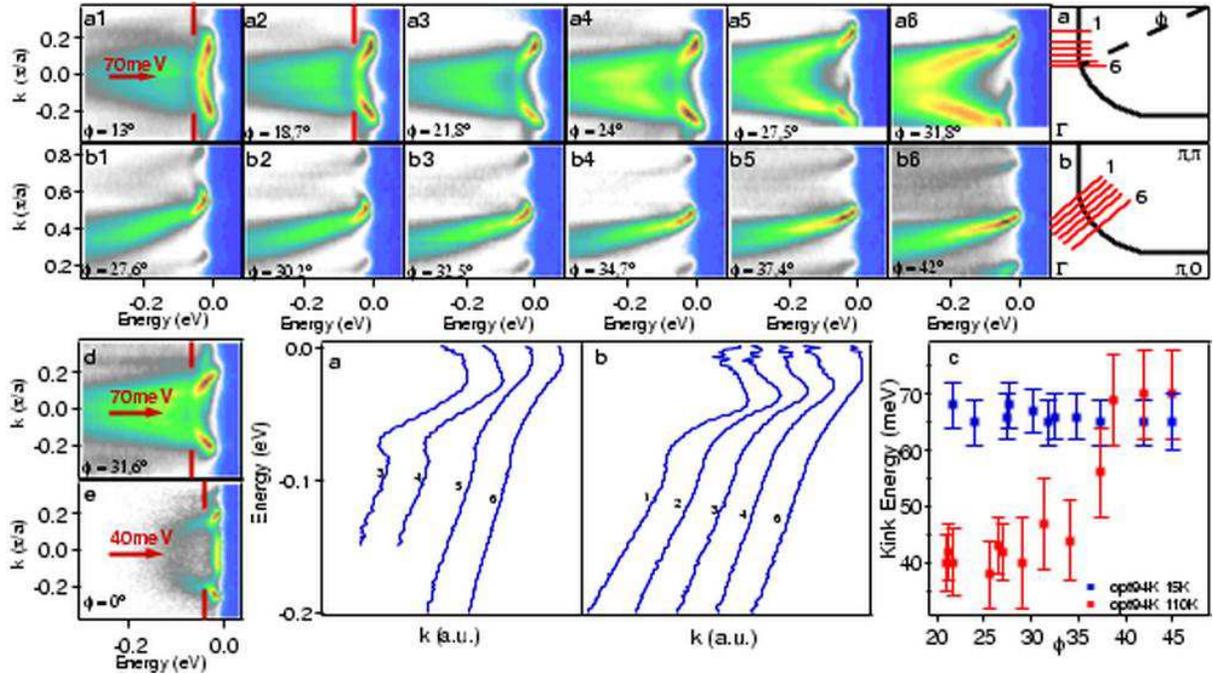}
\caption{\label{Figure1} 
The image plots in a1-a6), and the corresponding
MDC-derived dispersions, are cuts taken parallel to $(0,\pi)$-
$(\pi,\pi)$ at the locations indicated in the sketched zone at 15K.
The image plots in b1-b6), and the corresponding MDC-derived
dispersions, are cuts taken parallel to $(0,0)-(\pi,\pi)$, at
k-space locations indicated in the zone at 15K. Kink energies
as a function of angle $\phi$ are summarized in c) for optimally
(94K) doped samples in the normal and superconducting
states. d) and e) are spectra taken parallel to $(0,\pi)-(\pi,\pi)$
at the k-space locations indicated for under-doped (85K) and
over-doped samples (65K) of Bi-2212, respectively.}
\end{figure*}

In this paper we focus attention on some recent developments
coming from angle-resolved photoemission (ARPES) experiments and
theoretical developments of anisotropic electron-phonon
interactions in general.

Initially, the attention to bosonic renormalization effects in
cuprate superconductors was focussed on a ``kink" in the
electronic dispersion near 50-70 meV for nodal
electrons\cite{kink1,kink2,kink3,kink1SR,kink2SR} or solely below
T$_{c}$\cite{kink1SR,Kim,Sato,Gromko}. This has been further shown
to be generic to many hole-doped cuprates and is seen both in the
normal and superconducting states\cite{Tanja}. Recently, the
``kink" phenomenon has been reported for electronic states
throughout the Brillouin zone (BZ) again both above and below
T$_{c}$\cite{Tanja}. Although many-body effects reminiscent of strong coupling
are known to exist below T$_{c}$ in an extended k-space range, the observation
of kinks in the normal state imposes an additional constraint.
In contrast to the nodal renormalization
which shows little change across T$_c$, data in the
anti-nodal region reveal a dramatic change in the effective
coupling through T$_{c}$\cite{Tanja}, as shown in Fig.
\ref{Figure1} for 
Bi$_{2}$Sr$_{2}$Ca$_{0.92}$Y$_{0.08}$Cu$_{2}$O$_{8+\delta}$ (Bi-2212). 
In the superconducting state, classical
Engelsberg-Schrieffer signatures of electronic coupling to a
bosonic mode are seen in the image plot. The image plot shows
strong "kinks" or breaks in the energy dispersion of the band near
the $(\pi,0)$ regions of the BZ, and weaker kinks near the nodal
directions. 
The shift in the energy at which the 40meV mode couples
to the electrons ($\phi < 30$ deg) compounded by the lack
of shift in the $\sim 60-70$ meV nodal kink ($\phi >$ 35 deg) gives
rise to a fairly uniform $\sim 65-70$ meV kink energy throughout
the BZ in the superconducting state. The 
coupling strength, on the other hand, has a strong momentum
dependence below T$_{c}$. The relative sharpness
of the Momentum Distribution Curves (MDC)
kink (Fig. 1(a,b))indicates that the effective
coupling strength increases from the node to the antinode
of the $d$-wave gap. One can also see the increase
in coupling in the image plots of the raw data towards
$(\pi,0)$ as there is a stronger depression in intensity at the
mode energy. When the coupling gets very strong as in
Fig. 1(a1,a2), little dispersion can be tracked, and this is
the primary indicator of coupling.
The kinks occur for different dopings, and are clearly
seen in underdoped as well as overdoped samples, as shown in
Figure \ref{Figure1}d,e, respectively. In the deeply overdoped
sample (T $_{c}$ $\sim$ 65K or $\delta \sim$ 22$\%$), Fig. 1e, the
kink energy moves to 40-45meV since $\Delta_{0}$ ($\sim$ 10-15meV)
becomes much smaller. In the underdoped sample (T $_{c}$ $\sim$
85K), Fig. 1d, the kink energy remains around 70meV since it has a similar
gap as the optimally doped sample ($\Delta_{0}$ $\sim$ 35-40meV).
The signatures of coupling remain strong throughout, although they
do noticeably increase from the overdoped to underdoped sample
when comparing data taken at the same $\phi$.

\begin{figure}[htb]
\includegraphics[width=0.5\textwidth]{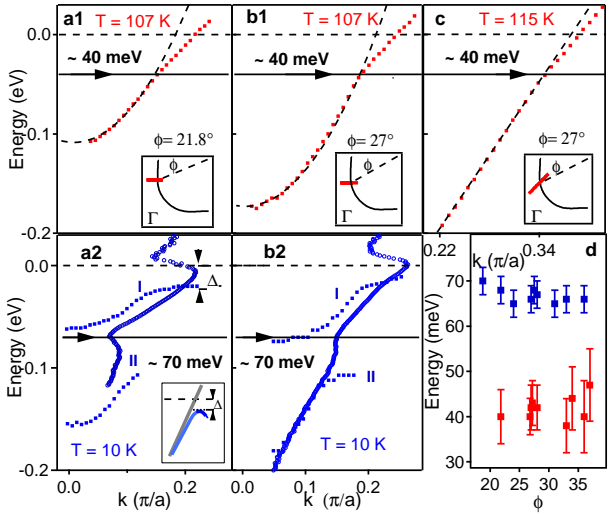}
\caption{\label{Figure2} EDC(a1, b1, c) derived dispersions in the
normal state (107K and 115K). $\phi$ and the cut-direction are
noted in the insets.  The red dots are the data; the fit to the
curve (dashes, black) below the 40meV line is a guide to the eye.
a2) and b2) are MDC derived dispersions at the same location and
direction as in a1) and b1), but in the superconducting state
(15K). In a2) and b2) we also plot the peak(I) and hump
positions(II) of the EDCs for comparison. The inset of a2) shows
the expected behavior of a Bogoliubov type gap opening. The s-like
shape below the gap energy is an artifact of how the MDC handles
the back-bend of the Bogoliubov quasiparticle. d) kink positions
as a function of $\phi$ in the anti-nodal region.}
\end{figure}

In the normal state little of these effects can be seen without
further analysis. As a consequence, kinks in the normal state have
been missed in prior works. In Fig. 2, we have extracted
dispersions for three k-space cuts in a momentum space region
between the nodal direction and the Van Hove Singularity (VHS) at
($\pi$,0). Since we are only concerned with the energy scale, we
fit the energy dispersion curves (EDCs) phenomenologically. The
usual method to extract dispersions---fitting MDC with Lorentzians---is not appropriate
since the assumed linear approximation of the bare band fails
towards ($\pi$,0) where the band bottom is close to
E$_{F}$\cite{LaShell}. EDC-derived dispersions of three
independent data sets shown in Fig. \ref{Figure2} consistently
reveal a $\sim$40meV energy scale in the normal state (a1,b1,c),
evolving into a 70 meV feature in the superconducting state
(a2,b2), as summarized in Figure \ref{Figure2}d. In the
superconducting state where the coupling is strongest, the peak
position in both EDC and MDC dispersions asymptotically approach
the characteristic energy defined by the bosonic mode (Fig. 2a2,
Fig. 2b2).

The kink energy shift is due to the opening of a superconducting
gap which shifts the energy at which the electronic states couple
to the bosonic mode. Here, we observe a kink shift of
$\sim$25-30meV, close to the maximum gap energy, $\triangle_{0}$.
We summarize the temperature dependence of the energy at which we
see a bosonic mode couple to the electronic states in the
anti-nodal region in Figs. \ref{Figure1}(c) and
\ref{Figure2}(d): the kink energies are
at $\sim$40meV above T$_{c}$ near the anti-nodal region and
increase to $\sim$70meV below T$_{c}$. Because the band minimum is
too close to E$_{F}$, the normal state $\sim$40meV kink cannot
easily be seen below $\phi\sim20^{o}$.

By comparing the different cut directions, one can clearly see
that the coupling is extended in the Brillouin zone and has a
similar energy scale throughout, near $\sim$ 70meV. Moreover, the
signatures of coupling increase significantly toward $(\pi,0)$ or
smaller $\phi$. The clear minimum in spectral weight seen near
70meV in Fig. 2a1) and 2a2) indicates strong mixing of the
electronic states with the bosonic mode where the two bare
dispersions coincide in energy.

The above described experimental observations are difficult to
reconcile in the spin resonance mode scenario. The kink is clearly
observed at 40meV in the normal state at optimal doping while the
41meV spin resonance mode exists only below
T$_{c}$\cite{Fong}. The kink is sharp in the superconducting
state of a deeply over-doped Bi2212 sample (Fig. 1e, consistent
with data of \cite{Gromko}) while no spin resonance mode has been
reported, or is expected to exist at this doping. In addition the
kink seen in under-doped Bi2212 (Fig. 1d) is just as sharp as in
the optimally doped case, while the neutron resonance peak is much
broader\cite{Fong}. Lastly, it is difficult to account for the
strong kink effect seen throughout the BZ from the spin mode since
the mode's spectral weight is only 2$\%$ \cite{Kee} as it may only
cause a strong enough kink effect if the interaction with
electrons is highly concentrated in k-space \cite{Abanov}.

On the other hand, it has been argued that the bosonic
renormalization effect may be reinterpreted as a result of a
coupling to the half-breathing in-plane Cu-O bond-stretching
phonon for nodal directions\cite{kink2,kink3} and the out-of-plane
out-of-phase O buckling $B_{1g}$ phonon for anti-nodal
directions\cite{Tanja}. This re-interpretation has the clear
advantage over the spin resonance in that it naturally explains
why the band renormalization effect is observed under three
circumstances: 1) in materials where no spin mode has been
detected, 2) in the normal state, and 3)in the deeply overdoped
region where the spin mode is neither expected nor observed.
However this interpretation also requires that the electron-phonon
interaction be highly anisotropic and its impact on the
electrons be strongly enhanced in the superconducting state.
This is something one does not expect a priori.

We now show that the anisotropy of the electron-phonon interaction
of these phonons, when formulated within the same framework as
that used to explain phonon lineshape changes observed via Raman
and neutron
measurements\cite{B1g,breathing,buckling,new}, explain
the observed momentum and temperature dependence of the data,
leading to unified understanding of band renormalizations. The
electron-phonon interactions are very anisotropic as a consequence
of the following four properties of the cuprates: 1) the symmetry
of the phonon polarizations and electronic orbitals involved
leading to highly anisotropic electron-phonon coupling matrix
elements $g(k,q)$; 2) the kinematic constraint related to the
anisotropy of the electronic band structure and the van Hove
singularity (VHS); 3) the d-wave superconducting gap, and 4) the
near degeneracy of energy scales of the $B_{1g}$ phonon, the
superconducting gap, and the VHS.

We consider a tight-binding three-band Hamiltonian modified by
$B_{1g}$ buckling vibrations as considered in Ref.
\cite{buckling} and in-plane breathing vibrations as considered
in Ref. \cite{Piekarz}:
\begin{eqnarray} \label{Hamiltonian}
&&H=\sum_{\textbf{n},\sigma}\epsilon_{Cu}b^{\dagger}_{\textbf{n},\sigma}b_{\textbf{n},\sigma}\nonumber
\\
&&+\sum_{\textbf{n},\delta,\sigma}\{\epsilon_{O}+eE_{z}u^{O}_{\delta}(\textbf{n})\}
a^{\dagger}_{\textbf{n},\delta,\sigma} a_{\textbf{n},\delta,\sigma}\nonumber\\
&&+\sum_{\textbf{n},\delta,\sigma}P_{\delta}[t-Q_{\delta}g_{dp}u^{O}_{\delta}(\textbf{n})]
(b_{\textbf{n},\sigma}^{\dagger}a_{\textbf{n},\delta,\sigma}
+ h.c.)\nonumber\\
&&+t^{\prime}\sum_{\textbf{n},\delta,\delta^{\prime}}P_{\delta,\delta^{\prime}}^{\prime}
(a_{\textbf{n},\delta}^{\dagger}a_{\textbf{n},\delta^{\prime},\sigma}
+ h.c)
\end{eqnarray}
\begin{figure}[htb]
\includegraphics[width= 5cm,height= 5cm]{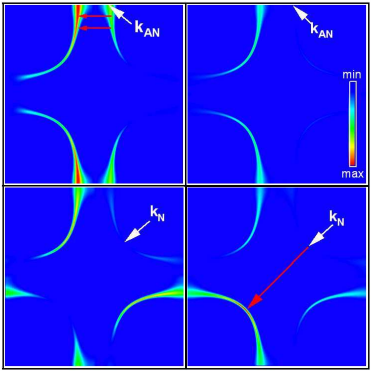}
\caption{\label{fig: coupling_constant} Plots of the
electron-phonon coupling $\mid g(\textbf{k},\textbf{q})\mid^{2}$
for initial $\textbf{k}$ and scattered $\bf{k^{\prime}=k-q}$
states on the Fermi surface for the buckling mode (left panels)
and breathing mode (right panels) for initial fermion $\textbf{k}$
at an anti-nodal (top panels) and nodal (bottom panels) point on
the Fermi surface, as indicated by the arrows. The red/blue color
indicates the maximum/minimum of the el-ph coupling vertex in the
BZ for each phonon.}
\end{figure}
with Cu-O hopping amplitude $t$, O-O amplitude $t^{\prime}$, and
$\epsilon_{d,p}$ denoting the Cu and O site energies,
respectively. Here $a_{\alpha=x,y},a^{\dagger}_{\alpha=x,y}$
($b,b^{\dagger}$) annihilates, creates an electron in the
O$_{\alpha}$ (Cu) orbital, respectively. The overlap factors
$Q_{\delta}, P_{\delta}, $ and $P_{\delta,\delta^{\prime}}^{\prime}$ are
given in our notation as $Q_{\pm x}=Q_{\pm y}=\pm 1; P_{\pm
x}=-P_{\pm y}=\pm 1; P_{\pm x,\pm y}^{\prime}=1=-P_{\mp x,\pm y}^{\prime}$. For the
half-breathing mode we consider only couplings $g_{dp}=q_{0}t$
(with $1/q_{0}$ a length scale) arising from modulation of the
covalent hopping amplitude $t$\cite{Piekarz,footnote}. For the
$B_{1g}$ mode a coupling to linear order in the atomic
displacements arises from a local c-axis oriented crystal field
$E_{z}$ which breaks the mirror plane symmetry of the Cu-O
plane\cite{buckling}. Such a field may be due to static buckling
or different valences of ions on either side of the Cu-O plane, 
orthorhombic tilts of the octahedra, or may be generated dynamically.

The amplitudes $\phi$ of the tight-binding wavefunctions forming
the anti-bonding band with energy dispersion $\epsilon({\bf k})$
arising from Eq. \ref{Hamiltonian} include the symmetry of the
$p,d$ wavefunctions of the Cu-O plane. The specific transformation
is constructed from the projected part of the tight-binding
wavefunctions
$b_{\textbf{k},\sigma}=\phi_{b}(\textbf{k})d_{\textbf{k},\sigma}$
and
$a_{\alpha,k,\sigma}=\phi_{\alpha}(\textbf{k})d_{\textbf{k},\sigma}$,
\begin{equation}
\label{wavefunction1}
\phi_{b}(\textbf{k})=\frac{1}{N(\textbf{k})}[\epsilon^{2}(\textbf{k})-t^{\prime
2}(\textbf{k})],
\end{equation}
\begin{equation}
\label{wavefunction2} \phi_{x,y}(\textbf{k})=\frac{\mp
i}{N(\textbf{k})}[\epsilon(\textbf{k})t_{x,y}(\textbf{k})-t^{\prime}(\textbf{k})t_{y,x}(\textbf{k})],
\end{equation}
with  $N^{2}(\textbf{k})=[\epsilon^{2}(\textbf{k})-t^{\prime
2}(\textbf{k})]^{2}+
[\epsilon(\textbf{k})t_{x}(\textbf{k})-t^{\prime
}(\textbf{k})t_{y}(\textbf{k})]^{2}+
[\epsilon(\textbf{k})t_{y}(\textbf{k})-t^{\prime
}(\textbf{k})t_{x}(\textbf{k})]^{2}$,
$t_{\alpha}(\textbf{k})=2t\sin(k_{\alpha}a/2),$ and
$t^{\prime}(\textbf{k})=4t^{\prime}\sin(k_{x}a/2)\sin(k_{y}a/2)$.
In particular if ${\bf k}$ resides on the Fermi surface, then
\begin{equation}
\label{wavefunction3} \phi_{b}^{FS}(\textbf{k})=\frac{-\mid
t^{\prime}(\textbf{k})\mid}{\sqrt{t^{\prime}(\textbf{k})^{2}+t_{x}(\textbf{k})^{2}+t_{y}(\textbf{k})^{2}}},
\end{equation}
\begin{equation}
\label{wavefunction4} \phi_{x,y}^{FS}(\textbf{k})=\frac{\pm i
sgn[t^{\prime}(\textbf{k})]t_{y,x}(\textbf{k})}{\sqrt{t^{\prime}(\textbf{k})^{2}+t_{x}(\textbf{k})^{2}+t_{y}(\textbf{k})^{2}}}.
\end{equation}

\begin{figure}[htb]
\includegraphics[width= 0.5\textwidth,clip=]{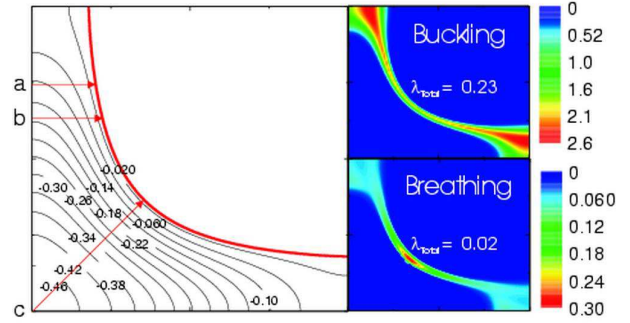}
\caption{\label{Fig:lambda} Plots of the electron-phonon coupling
$\lambda_{\bf k}$ in the first quadrant of the BZ for the buckling
mode (right top panel) and breathing mode (right bottom panel).
The color scale is shown on the right for each phonon. The left
panel shows energy contours for the band structure
used\cite{neutron}.}
\end{figure}

The full anisotropy of the electron-phonon coupling comes from the
projected wavefunctions in combination with the phonon
eigenvectors:
$H_{el-ph}=\frac{1}{\sqrt{N}}\sum_{\textbf{k},\textbf{q},\sigma,\mu}
g_{\mu}(\textbf{k},\textbf{q})c^{\dagger}_{\textbf{k},\sigma}c_{\textbf{k+q},\sigma}
[a^{\dagger}_{\mu,\textbf{q}}+a_{\mu,-\textbf{q}}]$. Neglecting
the motion of the heavier Cu atoms compared to O in the phonon
eigenvectors the form of the respective couplings can be compactly
written as
\begin{eqnarray}
\label{Eq:coupling_constant1} && g_{B_{1g}}({\bf k,q})= eE_{z}
\sqrt{\frac{\hbar}{4M_{O}M({\bf q})\Omega_{B_{1g}}}}
\\
&&\times\left\{\phi_{x}({\bf k})\phi_{x}({\bf
k^{\prime}})\cos(q_{y}a/2)-
\phi_{y}({\bf k})\phi_{y}({\bf k^{\prime}})\cos(q_{x}a/2)\right\}\nonumber\\
\label{Eq:coupling_constant2}&&g_{br}({\bf
k,q})=g_{dp}\sqrt{\frac{\hbar}{2M_{O}\Omega_{br}}}
\sum_{\alpha=x,y}\\
&&\times\left\{\phi_{b}({\bf
k^{\prime}})\phi_{\alpha}(\textbf{k})\cos(k^{\prime}_{\alpha}a/2)-
\phi_{b}(\textbf{k})\phi_{\alpha}({\bf
k^{\prime}})\cos(k_{\alpha}a/2)\right\} \nonumber,
\end{eqnarray}
with $M({\bf q})=[\cos^{2}(q_{x}a/2)+\cos^{2}(q_{y}a/2)]/2$ and
${\bf k^{\prime}}=\textbf{k-q}$. Here we have neglected an overall
irrelevant phase factor. The anisotropy can be shown at small
momentum transfers ${\bf q}$ where $g_{B_{1g}}({\bf k, q=0})\sim
\cos(k_{x}a)-\cos(k_{y}a)$, while $g_{br}\sim \sin(qa)$ for any
\textbf{k}. Also for large momentum transfer ${\bf
q}=(\pi/a,\pi/a)$, $g_{B_{1g}}$ vanishes for all ${\bf k}$ while
$g_{br}$ has its maximum for \textbf{k} located on the nodal
points of the Fermi surface. While this $q$-dependence has been
the focus before in context with a $d_{x^{2}-y^{2}}$ pairing
mechanism\cite{phonons}, the dependence on fermionic wavevector
${\bf k}$ has usually been overlooked. This strong momentum dependence
of the $B_{1g}$ coupling implies that resistivity measurements would
be sensitive only to the weaker part of the electron-phonon coupling, in
agreement with experiments. It is exactly this
fermionic dependence which we feel is crucially needed to
interpret ARPES data.

\begin{figure*}[htb]
\includegraphics[width= 1\textwidth]{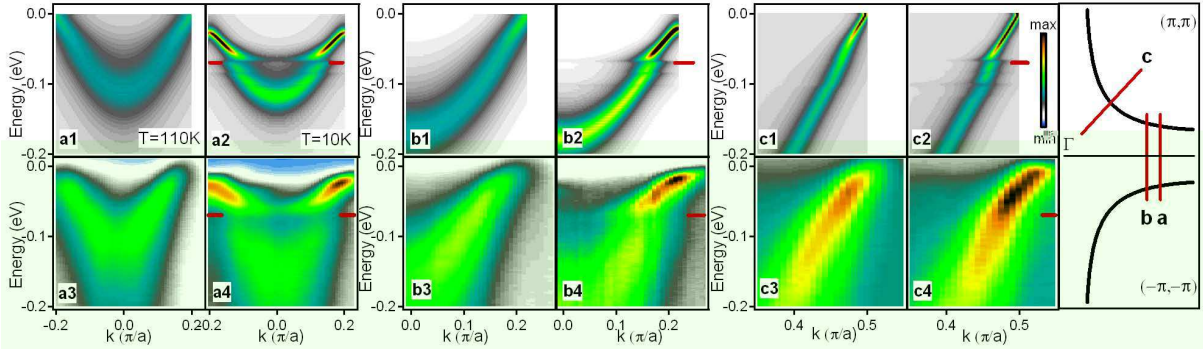}
\caption{\label{Fig:3} Image plots of the calculated spectral
functions in the normal (a1,b1,c1) and superconducting (a2,b2,c2)
states compared to the spectral functions in the normal (a3,b3,c3)
and superconducting (a4,b4,c4) states measured in
Bi$_{2}$Sr$_{2}$Ca$_{0.92}$Y$_{0.08}$Cu$_{2}$O$_{8+\delta}$
(Bi-2212)\cite{Tanja} for momentum cuts {\emph{a,b,c}} shown in
the right-most panel and in Figure 2. The same color scale is used
for the normal/superconducting pairs within each cut, but the
scaling for the data and the calculation are separate. The red
markers indicate 70meV in the superconducting state.}
\end{figure*}

Fig \ref{fig: coupling_constant} plots $\mid
g(\textbf{k},\textbf{q})\mid^{2}$ as a function of phonon
scattered momentum $\textbf{q}$ connecting initial $\textbf{k}$
and final $\bf{k^{\prime}=k-q}$ fermion states both on the Fermi
surface. Here we have adopted a five-parameter fit to the band
structure $\epsilon({\bf k})$ used previously for optimally doped
Bi-2212\cite{neutron}. For an anti-nodal initial fermion
$\textbf{k}_{AN}$, the coupling to the $B_{1g}$ phonon is largest
for $q=0$ and $2\textbf{k}_{F}$ scattering to an anti-nodal final
state. The coupling to the breathing mode is much weaker overall,
and is especially weak for small $\textbf{q}$ and mostly connects
to fermion final states at larger $\textbf{q}$. For a nodal
fermion initial state $\textbf{k}_{N}$, the $B_{1g}$ coupling is
suppressed for small $\textbf{q}$ and has a weak maximum towards
anti-nodal final states, while the breathing mode attains its
largest value for scattering to other nodal states, particularly
to state $-\textbf{k}_{N}$.

In Fig. \ref{Fig:lambda} we plot $\lambda_{\bf k}$ for each k
point in the BZ in the normal state, where we have
defined\cite{Jepsen}:
\begin{eqnarray}
&&\lambda_{\bf k}=\frac{2}{N_{F}\Omega^{2}_{0}N^{2}} \sum_{\bf q}
[f(\epsilon({\bf k}))-f(\epsilon({\bf k+q}))]\nonumber\\
&&\times \mid g({\bf k,q})\mid^{2}\delta(\epsilon({\bf
k+q})-\epsilon({\bf k})-\hbar\Omega_{0}),
\end{eqnarray}
with $N_{F}={1\over{N}}\sum_{\bf k}\delta(\epsilon({\bf k}))$ the
density of states per spin at the Fermi level. As a result of the
anisotropy of the coupling constant, large enhancements of the
coupling near the anti-node for the $B_{1g}$ phonon, and nodal
directions for the breathing phonon mode are observed,
respectively, which are an order of magnitude larger than 
$\lambda={1\over{N}}\sum_{\bf k}\lambda_{\bf k}$. Here we have used the
value of the electric field $eE_{z}=1.85 eV/\AA$ consistent with
Raman scattering measurements of the $B_{1g}$ phonon and buckling
of the Cu-O plane in YBaCuO, and have taken
$q_{0}t=2\sqrt{2}eE_{z}$ as a representative value for the
coupling to the breathing mode. These values lead to net couplings
similar to that obtained from LDA calculations for an infinite
layer compound CaCuO$_{2}$ which develops a static
dimple\cite{Jepsen}.

The Nambu-Eliashberg electron-phonon self-energy determines
bosonic features in the spectral function\cite{Sandvik}:
$\hat\Sigma(\textbf{k},i\omega_{m})=
i\omega_{m}(1-Z(\textbf{k},i\omega_{m}))\hat\tau_{0} +
\chi(\textbf{k},i\omega_{m})\hat\tau_{3}+\Phi(\textbf{k},i\omega_{m})\hat\tau_{1}$
with
\begin{eqnarray}
\label{Eliashberg} &&\hat\Sigma(\textbf{k},i\omega_{m})=
\frac{1}{\beta N}\sum_{\textbf{p},\mu,m}g_{\mu}(\textbf{k,p-k})
g_{\mu}(\textbf{p,k-p})\nonumber\\
&&\times D^{0}_{\mu}(\textbf{k-p},i\omega_{m}-i\omega_{n})
\hat\tau_{3}\hat G^{0}(\textbf{p},i\omega_{m})\hat\tau_{3},
\end{eqnarray}
with $\hat G^{0}(\textbf{p},i\omega_{m})=
\frac{i\omega_{m}\hat\tau_{0}+\epsilon(\textbf{p})\hat\tau_{3}+\Delta(\textbf{p})\hat\tau_{1}}{(i\omega_{m})^{2}-
E^{2}(\textbf{p})}$ and
$E^{2}(\textbf{k})=\epsilon^{2}(\textbf{k})+\Delta^{2}(\textbf{k})$,
with $\hat\tau_{i=0\cdots 3}$ are Pauli matrices. We take
$\Delta(\textbf{k})=\Delta_{0}[\cos(k_{x}a)-\cos(k_{y}a)]/2$ with
$\Delta_{0}=35$meV. The bare phonon propagators are taken as
$D^{0}_{\mu}(\textbf{q},i\Omega_{\nu})=\frac{1}{2}\left[
\frac{1}{i\Omega_{\nu}-\Omega_{\mu}}-
\frac{1}{i\Omega_{\nu}+\Omega_{\mu}}\right].$ Here we take
dispersionless optical phonons $\hbar\Omega_{B_{1g},br}=36,70$meV,
respectively.

The spectral function $A(\textbf{k},\omega)=-\frac{1}{\pi}Im
G_{11}(k,i\omega\rightarrow\omega+i\delta)$ is determined from Eq.
\ref{Eliashberg} with $\hat
G(k,i\omega_{m})=[\hat\tau_{0}i\omega_{m}Z(\textbf{k},i\omega_{m})-(\epsilon(\textbf{k})+
\chi(\textbf{k},i\omega_{m}))\hat\tau_{3}-\Phi(\textbf{k},i\omega_{m})\hat\tau_{1}]^{-1}$
and the coupling constants given by Eq.
\ref{Eq:coupling_constant1}-\ref{Eq:coupling_constant2}. Here we
consider three different cuts as indicated in Fig.
\ref{Fig:lambda}: along the nodal direction ($k_{x}=k_{y}$,
denoted by \emph{c}), and two cuts parallel to the BZ face
($k_{y}=0.75\pi/a$, \emph{b}) and ($k_{y}=0.64\pi/a$, \emph{a})
where bilayer effects are not severe\cite{Tanja}.

Our results for the three cut directions are shown in Fig
\ref{Fig:3} for the normal ($T=110K$) and superconducting
($T=10K$) states. Very weak kink features are seen in the normal
state, with a smeared onset of broadening at the phonon energy due to the measurement
temperature. A
much more pronounced kink at 70 meV in the superconducting state
is observed for all the cuts along with a sharp onset of
broadening, in excellent agreement with the data\cite{Tanja},
shown also in Fig. \ref{Fig:3}. In the superconducting state, the
kink sharpens due to the singularity in the superconducting
density of states from the anti-nodal regions. Thus cut $a$
exhibits the strongest signatures of coupling, showing an Einstein
like break up into a band that follows the phonon dispersion, and
one that follows the electronic one. In cut $b$, a weaker
signature of coupling manifests itself in an $s-$like
renormalization of the bare band that traces the real part of the
phonon self-energy, while in cut $c$, where the coupling to the
$B_{1g}$ phonon is weakest, the electron-phonon coupling manifests
itself as a change in the velocity of the band. The agreement is
very good, even for the large change in the electron-phonon
coupling seen between cuts $a$ and $b$, separated by only $1/10$th
of the BZ.

We now discuss the reason for the strong anisotropy as a result of
a concurrence of symmetries, band structure, $d-$wave energy gap,
and energy scales in the electron-phonon problem. In the normal
state, weak kinks are observed at both phonon frequencies, but
more pronounced at 70 meV for cut \emph{c} and at 36 meV for cut
\emph{a}, as expected from the anisotropic couplings shown in Fig.
\ref{fig: coupling_constant}. However this is also a consequence
of the energy scales of the phonon modes and the bottom of
electronic band along the cut direction as shown in Fig.
\ref{Fig:lambda}. Further towards the anti-node, another
interesting effect occurs. As one considers cuts closer to the BZ
axis, the bottom of the band along the cut rises in energy and the
breathing phonon mode lies below the bottom of the band for
($k_{x}=0,k_{y}>0.82\pi/a$) and so the kink feature for this mode
disappears. This is the usual effect of a Fano redistribution of
spectral weight when a discrete excitation lies either within or
outside of the band continuum\cite{Fano}. This thus can be used to
open "windows" to examine the coupling of electrons to discrete
modes in general. Taken together, this naturally explains why the
normal state nodal kink is near 70
meV\cite{kink1,kink2,kink3,kink1SR,kink2SR} while the anti-nodal
kink is near 40 meV\cite{Kim,Sato,Gromko,Tanja}.

In the superconducting state the bare band renormalizes downward
in energy due to the opening of the gap and gives the strongest
effect for the anti-nodal region where the gap is of the order of
the saddle point energy. The $B_{1g}$ coupling becomes dramatically
enhanced by the opening of the gap at a frequency close to the
frequency of the phonon itself, and leads to the dramatic
renormalization of the phonon observed in Raman and neutron
measurements\cite{B1g}. Yet the breathing coupling does not vary
dramatically as the large gap region is not weighted as heavily by
the coupling constant. As a consequence, the $B_{1g}$ mode
dominates and clear kinks in the data are revealed at around 70
meV - the $B_{1g}$ energy plus the gap. Indeed a much weaker kink
at 105 meV from the breathing phonon shows at higher energies for
the nodal cut but this kink becomes weaker away from nodal
directions as the coupling and the band bottom along the cut moves
below the phonon frequency so that the ``energy window'' closes.

\begin{figure}[htb]
\includegraphics[width=0.5\textwidth]{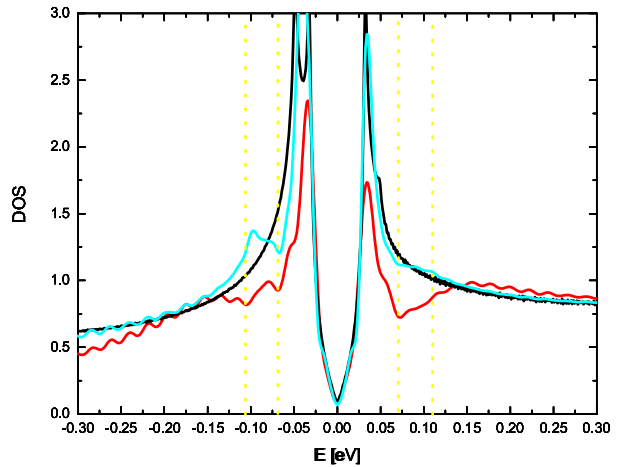}
\caption{\label{Fig:6} Superconducting density of states
calculated with the electron-phonon self energy for three
different values of the electric field $eE_{z}$=0 (black), 1.85
 (blue), and 3.2 (red), in units of $eV/\AA$,
respectively. The yellow dashed lines indicate the energies
$\Omega_{B1g}+\Delta_{0}$ and $\Omega_{br}+\Delta_{0}$.}
\end{figure}

Finally we plot in Fig. \ref{Fig:6} the superconducting density of
states (DOS) $-1/\pi Im G_{11}(k,\omega)$ calculated with the
electron-phonon contributions from both the breathing and $B_{1g}$
phonons to the self energy determined via Eq. 9 for three
different values of the $B_{1g}$ coupling constant parametrized by
the crystal field $eE_{z}$. We have fixed the relative breathing
coupling at $q_{0}t=2\sqrt{2}eE_{z}$ as for ARPES. For these
values of the coupling constants the structure in the DOS is
determined largely by the anti-nodal couplings to the $B_{1g}$
phonon. For the value of the coupling which fits the ARPES data
well, the peak-dip-hump structure of the DOS (at values
$E=\Delta_{0},\Omega_{B1g}+\Delta_{0},$ and $\Omega_{B1g}+E_{vH}$,
respectively) compares very well with the DOS determined via
scanning tunnelling microscopy in Bi-2212\cite{STM}.

In summary, contrary to usual opinion of the role of anisotropy in
electron-phonon coupling, the interplay of specific coupling
mechanisms of electrons to the buckling and breathing phonons give
a natural interpretation to the bosonic renormalization effects
seen in ARPES in both the normal and superconducting states, and
provides a framework to understand renormalizations as a function
of doping.

\textit{Acknowledgments}: ARPES data were collected at SSRL which
is operated by DOE under contract DE-AC03-76SF00515. T.P.D. would
like to acknowledge support from NSERC, PREA and the A. von
Humboldt Foundation. The Stanford work is also supported by NSF
grant DMR-0304981 and ONR grant N00014-01-1-0048.


\begin{thebibliography}{00}
\bibitem{footnote}
The coupling $g_{d}$ arising from modulation of Cu site energies
leads to a coupling which only depends on momentum transfers on
not on fermion momentum\cite{Piekarz}.
\end{thebibliography}
\end{document}